\documentclass[11pt]{article}
\usepackage{jheppub}
\usepackage{graphicx}
\usepackage{amsmath,amssymb}
\usepackage{array}
\usepackage{comment}
\usepackage{hyperref}
\usepackage{pb-diagram}
\usepackage{slashed}
\usepackage{cancel}
\usepackage{float}
\usepackage[normalem]{ulem}
\usepackage{tcolorbox}
\usepackage{epsfig,amsfonts}
\usepackage{mathrsfs}
\usepackage[title]{appendix}
\usepackage{mathtools}
\usepackage{physics}

\usepackage[utf8]{inputenc}

\usepackage{tikz}
\usetikzlibrary{shapes.geometric, arrows}
\tikzstyle{decision} = [ellipse, minimum width=3cm, minimum height=1cm, text centered, draw=black, text width=2cm, fill=green!30]
\tikzstyle{rect} = [rectangle, rounded corners, minimum width=1cm, minimum height=1cm,text centered, fill=yellow, draw=black]
\tikzstyle{trap} = [trapezium, rounded corners,minimum width=1cm, centered, fill=orange, draw=black]
\tikzstyle{arrow} = [thick,->,>=stealth]

\newcommand{\be}{\begin{equation}}
\newcommand{\ee}{\end{equation}}
\newcommand{\bea}{\begin{eqnarray}}
\newcommand{\eea}{\end{eqnarray}}

\newcommand{\Mpl}{M_\mathrm{pl}}

\newcommand{\At}{\widetilde{A}}
\newcommand{\gt}{\widetilde{g}}
\newcommand{\Vt}{\widetilde{V}}
\newcommand{\Rt}{\widetilde{R}}

\usepackage{color}
\usepackage{xcolor}

\begin{document} 

\title{\hfill ~ \\[-40mm]
\begin{footnotesize}
\hspace{126mm}
\normalfont{DIAS-STP-25-07}\\
\end{footnotesize}
\vspace{30mm}
\boldmath
Chaotic Inflation RIDES Again}
\author[\,a,b]{Venus Keus}
\author[\,c]{and Stephen F. King}

\affiliation[a]{School of Theoretical Physics, Dublin Institute for Advanced Studies, 10 Burlington Road, Dublin, D04 C932, Ireland}
\affiliation[b]{Department of Physics and Helsinki Institute of Physics, Gustaf Hallstromin katu 2, FIN-00014, University of Helsinki, Finland}
\affiliation[c]{School of Physics and Astronomy, University of Southampton, Southampton, SO17 1BJ, United Kingdom}

\emailAdd{venus@stp.dias.ie}
\emailAdd{s.f.king@soton.ac.uk}

\abstract{
Following the recent Atacama Cosmology Telescope (ACT) results, we revisit 
chaotic inflation based on a single complex scalar field with mass term $M^2 |\Phi|^2$, which usually predicts a spectral index $n_s\approx 0.96$
but a too-large tensor to scalar ratio 
$r\approx 0.16$. With radiative corrections, the potential $M^2 |\Phi|^2 \ln \left( |\Phi|^2/\Lambda^2 \right)$ induces spontaneous symmetry breaking near the scale $\Lambda$, yielding a Pseudo Nambu-Goldstone boson which can play the role of a quintessence field, hence radiative inflation and dark energy (RIDE). Including a non-minimal coupling to gravity
 $\xi  |\Phi|^2 R^2$ with $\xi \sim 0.1$ reduces the tensor to scalar ratio to $r \lesssim0.03$, allowing a good fit of the RIDE model to Planck data. Allowing a small additional quartic coupling correction $\lambda |\Phi|^4$ allows a good fit to ACT data sets for $\xi \sim 1$ and $\lambda \sim 10^{-5}$.
\\ \\ \\ \today
}

\maketitle

\section{\label{sec:Introduction}Introduction}

Dark energy, the simplest example being Einstein's cosmological constant $\Lambda$, together with cold dark matter (CDM) form part of the standard cosmological model, often referred to as $\Lambda$CDM
(for a pedagogical introduction see e.g.~\cite{DiBari:2018vba}). Despite its many successes, $\Lambda$CDM provides no understanding of either dark matter or dark energy, and also leaves many underlying theoretical questions unanswered (see e.g.~\cite{Dimopoulos:2020pjx}).

Cosmic inflation \cite{Guth:1980zm} remains an important extension of the $\Lambda$CDM, capable of explaining  the flatness and homogeneity of
the universe as well as adiabatic perturbations by postulating 
an exponential expansion of space at very early times.
Slow-roll inflation \cite{Linde:1981mu,Albrecht:1982wi}, with a slowly rolling scalar inflaton field,
predicts an approximately scale invariant and Gaussian spectrum \cite{Starobinsky:1980te}, which has been confirmed by
observations of the large scale structure of the universe
and the cosmic microwave background (CMB) \cite{Martin:2013tda}. 

Chaotic inflation is one of the earliest and simplest form of slow roll inflation since, avoiding the initial condition problem by allowing the scalar field to take some random (chaotic) large initial value.
The simplest example of chaotic inflation, based on the mass term 
$M^2 \phi^2$, where $\phi$ is a real scalar field, and $M$ is its mass~\cite{Linde:1983gd}, predicts a tensor to scalar ratio of around 0.16 which is too large to account for current observations. 
This is unfortunate, because it would permit a very simple interpretation in terms of right-handed sneutrinos (see e.g. \cite{Ellis:2004hy} and references therein).
However, if the scalar field couples to gravity with a non-minimal coupling
$\xi  \phi^2 R^2$, where $R$ is the Ricci scalar, then consistency with data may in principle be achieved for some non-minimal coupling $\xi$~\cite{Linde:2011nh,Kallosh:2025rni}, as discussed recently as part of a general study~\cite{Kodama:2021yrm}.

In this paper, we consider chaotic inflation with a complex scalar field $\Phi$, and include the effect of radiative corrections, via the phenomenological potential 
$M^2 |\Phi|^2 \ln \left( |\Phi|^2/\Lambda^2 \right)$~\cite{DiBari:2010wg,DiBari:2014oja}.
A notable feature of having a complex scalar field with such radiative corrections is to change the shape of the potential to a Mexican hat type of potential. This leads to chaotic inflation for large values of $|\Phi|$, together with spontaneous symmetry breaking at the minimum of the potential, resulting in a pseudo Nambu-Goldstone boson (PNGB), which can then be used as a quintessence field.~\footnote{The quintessence potential generated from gravitational effects was discussed in general terms in~\cite{Kallosh:1995hi}.}
The resulting scheme was dubbed radiative inflation and dark energy (RIDE)~\cite{DiBari:2010wg,DiBari:2014oja}.
The RIDE model is particularly attractive since both inflation and dark energy energy emerge from a single complex scalar field $\Phi$.
The radiatively corrected potential studied in~\cite{DiBari:2010wg,DiBari:2014oja}, although consistent with the CMB data available at the time (WMAP 7-year), is in tension with the Planck 2018 constraints~\cite{Planck:2018jri}.  In this work we show that introducing a non-minimal coupling to gravity, $\xi |\Phi|^2 R^2$, allows the model to achieve good agreement with the Planck data. To further improve the fit to the recent ACT results~\cite{ACT:2025tim}, we also include a small quartic interaction $\lambda |\Phi|^4$ and perform a detailed analysis of the parameter space in terms of $\xi$ and $\lambda$.
Other recent studies that introduce related modifications to various inflationary potentials to accommodate the ACT data include~\cite{Yin:2025rrs,Gialamas:2025kef,Liu:2025qca,Haque:2025uis,Han:2025cwk,Zharov:2025zjg,Pallis:2025vxo,Yuennan:2025inm,Addazi:2025agg,Kumar:2025apf}.
In the present paper, these corrections are applied to one of the simplest and earliest inflationary potentials, namely that of chaotic inflation, including radiative corrections, with the added feature that it can also accommodate dark energy. 

The layout of the remainder of the paper is as follows.
In section~\ref{sec:original-RIDE} we review the original RIDE model, and its predictions for inflation and quintessence, comparing its predictions to recent data.
In section~\ref{sec:improved-RIDE} we show how the prospects for the RIDE model may be considerably improved by including a non-minimal coupling to gravity, which gives consistency with recent Planck data, and an optional small quartic coupling in order to allow a better fit to recent ACT results. Appendix~\ref{sec:Appendix-A} details the conformal transformation from the Jordan frame to the Einstein frame. 
Appendix~\ref{sec:Appendix-B} includes the definition of data sets used in the analysis of the ACT collaboration.

\section{\label{sec:original-RIDE}The original RIDE model}

\subsection{The model}

The starting point of the model is
the simple chaotic inflation mass term, but involving a complex scalar field $\Phi$, which as usual is a gauge singlet.
In the absence of radiative corrections the potential has a simple quadratic form
\be 
V_0\approx M^2 |\Phi|^2 . 
\ee
As in chaotic inflation, the quartic coupling is assumed to be negligibly small. 
\footnote{For example, the potential could arise from a supersymmetric Wess-Zumino model with a superpotential 
$W=M \Phi^2$ where cubic terms are forbidden by a discrete symmetry and non-renormalisable quartic terms are suppressed by some high scale.
}
The effect of introducing quartic couplings in the potential is studied in Section~\ref{sec:original-RIDE+quartic}.

The RIDE model assumes that radiative corrections,
arising from some unspecified Planck scale interactions, lead to a modified potential which may be parameterised as~\cite{DiBari:2010wg,DiBari:2014oja},
\be 
V \, \approx \, M^2 |\Phi|^2 \,  \ln \left( \frac{|\Phi|^2}{\Lambda^2} \right) 
\,
\label{eq:inf_pot-new}
\ee
leading to spontaneous symmetry breaking which occurs around the scale $\Lambda$.
Examples of models with radiative symmetry breaking can be found in Refs.~\cite{Ibanez:1982fr,deMedeirosVarzielas:2006gtm,Howl:2009ds,Stewart:1996ey,Stewart:1997wg,Covi:2004tp} where the focus is on the corrections due to the renormalization group evolution which is the dominant contribution of the Coleman-Weinberg correction~\cite{Coleman:1973jx} in the case of broken supersymmetry~\cite{Lyth:2007qh}. See also Ref.~\cite{Shafi:2006cs} for experimental constraints on such corrections. 
\footnote{The radiatively generated spontaneous symmetry breaking can also be interpreted as dynamical symmetry breaking, due to the  effective mass squared of the scalar field being driven negative at ``low'' energies, although in practice this may be at a scale $\Lambda$, not too far below the Planck scale.
 This mechanism of radiative symmetry breaking is well-known in the minimal supersymmetric standard model~\cite{Ibanez:1982fr}, where the Higgs mass squared is driven negative at the TeV scale. Similarly, radiative symmetry breaking can play an important role in different contexts~\cite{deMedeirosVarzielas:2006gtm,Howl:2009ds}, where a mass-squared is driven negative at a much higher scale. } 

The complex scalar field $\Phi$ has two components which may be parameterised in terms of the radial field 
$\sigma$ and the angular field $\mathrm{\varphi}$,
\be 
\Phi=\frac{1}{\sqrt{2}}\, \sigma \, e^{i\varphi/v_\sigma} \, .
\ee
The inflationary potential arises purely from the radial field and is of the form
\be 
\label{eq:V-st}
V(\sigma)=\frac{1}{2} M^2 \sigma^2 \ln \left( \frac{{\sigma}^2}{2\Lambda^2} \right)\,.
\ee
This leads to a vacuum expectation value (VEV) of $v_\sigma=\sqrt{\frac{2}{e}}\Lambda$ for $\sigma$, and inflation can take place in a region where $\sigma \gg \Lambda$, in which the $\ln$-term in Eq.~\eqref{eq:inf_pot-new} is well behaved and the inflaton field $\sigma$ only feels a potential that is very similar to the one used for quadratic inflation.

Later on, the field will settle at its VEV. As the potential is symmetric under a global $U(1)$, either imposed or accidental, the VEV will break this global symmetry, thereby generating a massless Nambu-Goldstone boson $\varphi=v_\sigma \arg (\Phi)$. This field has no mass term and in fact no potential at all. 
Gravitational effects, i.e. gravitational instantons, can break this symmetry and generate a tiny mass $m$~\cite{Kallosh:1995hi}.

The angular part of the complex scalar field $\varphi$, can play the role of a quintessence field, with a potential of the form~\footnote{Radiative corrections to the quintessence potential may also be important, see Refs.~\cite{Hill:1988bu,Frieman:1991tu}}
\be 
V(\varphi)=m^4 \left[ 1+\cos \left( \frac{\varphi}{v_\sigma} \right) \right] \, ,
 \label{eq:quint_pot-new}
\ee
where the value of the mass scale $m$ is assumed to be set by the cosmological constant~\cite{Kallosh:1995hi}.
We will discuss the resulting quintessence model in  Section~\ref{sec:Quintessence}.

Note that the dynamics of both sectors can be easily disentangled, as the kinetic term simplifies to
\be 
 (\partial_\mu \Phi^*) (\partial^\mu \Phi)=\frac{1}{2} (\partial_\mu \sigma)(\partial^\mu \sigma) + \frac{\sigma^2}{2 \,{v^2_\sigma}}(\partial_\mu \varphi) (\partial^\mu \varphi),
 \label{eq:kin_term}
\ee 
with the $\varphi$-part being negligible during inflation and $\sigma$ already sitting at its (constant) VEV $v_\sigma$ during quintessence. Due to this separation of the dynamics of the two fields, the model is basically a single-field inflationary model and is safe from iso-curvature fluctuations.

\begin{figure}[ht!]
\centering
\vspace{-4mm}
\includegraphics[height=78mm]{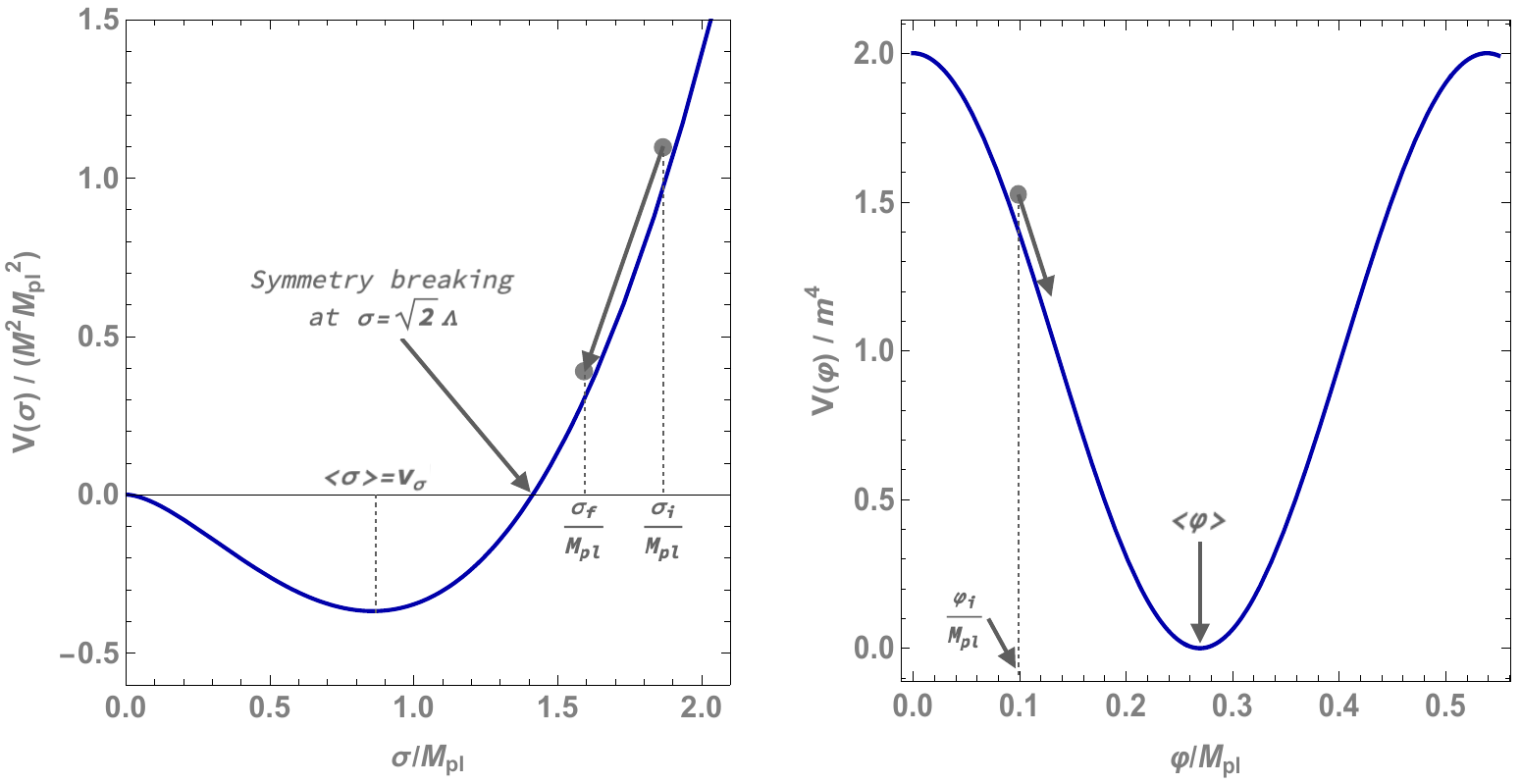}
\vspace{-4mm}
\caption{\label{fig:RIDE-potentials}The schematic shape of the (scaled) inflationary potential ${V(\sigma)}/{(M^2 \Mpl^2)}$ (left panel) and the (scaled) quintessence potential $ V(\varphi)/m^4$ (right panel) for the value of $\Lambda=\Mpl$.}
\end{figure}

\begin{figure}[ht!]
\centering
\includegraphics[height=47mm]{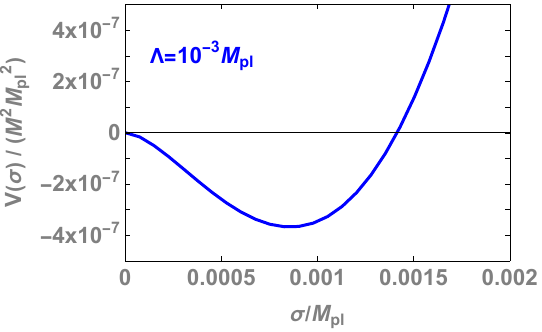}~~
\includegraphics[height=47mm]{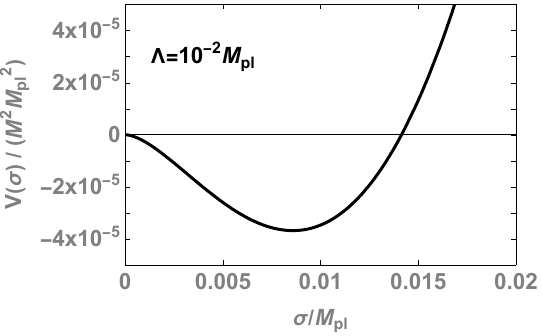}\\[2mm]
\includegraphics[height=47mm]{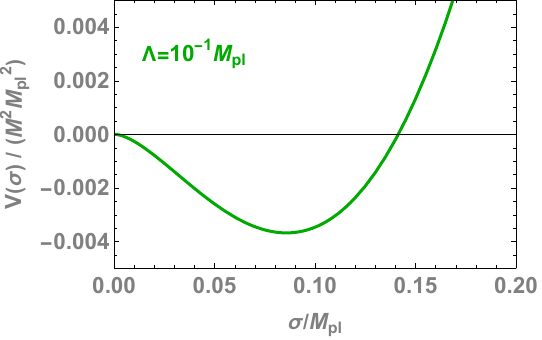}~~~~~~
\includegraphics[height=47mm]{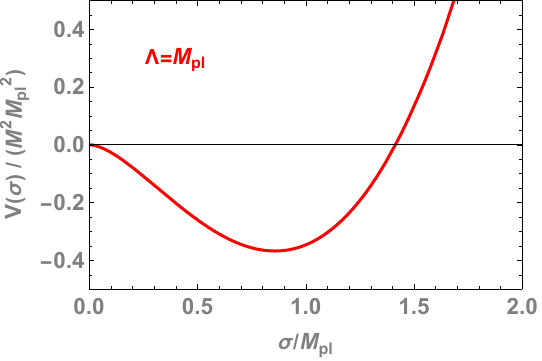}
\vspace{-3mm}
\caption{\label{fig:RIDE-potentials-k}The inflationary potential at different values of $\Lambda$. Note that its minimum occurs near the value of $\Lambda$ in each case.}
\end{figure}

In Fig.~\ref{fig:RIDE-potentials}, we show the 
inflationary potential in Eq.~\eqref{eq:V-st} scaled by the $(M^2 \Mpl^2)$ factor ${V(\sigma)}/{(M^2 \Mpl^2)}$ and the quintessence potential in Eq.~\eqref{eq:quint_pot-new} scaled by the $m^4$ factor $V(\varphi)/m^4$ for the value of $\Lambda=\Mpl$, where $\Mpl$ is the reduced Planck mass. 
In Fig.~\ref{fig:RIDE-potentials-k} we plot the inflationary potential for different values of $\Lambda$,
which controls the field value at the minimum.

\subsection{\label{sec:Inflation}Inflation in the RIDE model}
For completeness, in this section, we work out the slow-roll parameters of the original RIDE model, which are
\bea 
\epsilon_V &=& 
\frac{\Mpl^2}{16\pi} \left( \frac{V'}{V} \right)^2
=\frac{\Mpl^2}{4\pi \,\sigma^2} \left(1 + \frac{1}{\ln \left( \frac{\sigma^2}{2\Lambda^2} \right)} \right)^2
\\[2mm] 
\eta_V &=& 
\frac{\Mpl^2}{8\pi} \left( \frac{V''}{V} \right)
=\frac{\Mpl^2}{4\pi \,\sigma^2} \left( 1 + \frac{3}{\ln \left( \frac{\sigma^2}{2\Lambda^2} \right)}
\right)
 \label{eq:slow_roll}
\eea
During inflation, both parameters are assumed to satisfy $\epsilon, \eta \ll 1$ which satisfies the slow roll condition.

To calculate the values of $\sigma$ at the beginning and end of inflation, $\sigma_i$ and $\sigma_f$, respectively, one needs to calculate the number of e-folds $N_e$, i.e. the number of times the universe expanded by $e$ times its own size. $N_e$ is calculated to be
\bea 
N_e &=& \frac{8\pi}{\Mpl^2} \int_{{\sigma}_f}^{{\sigma}_i}\frac{V}{V^\prime} d{\sigma}
=
2\pi  \left[\frac{\sigma^2}{\Mpl^2} - \frac{2}{e}\left(\frac{\Lambda}{\Mpl}\right)^2  {\rm Ei}\left(1+\ln \left(\frac{{\sigma}^2}{2\Lambda^2}\right) \right) \right]
\label{eq:Ne}
\eea
where ${\rm Ei}(z)$ is the exponential integral ${\rm Ei}(z)=-\int_{-z}^{\infty} \frac{e^{-t}}{t} dt$. 
One can calculate ${\sigma}_f$ by setting $\epsilon=1$ at the end of inflation. 
We solve the $\epsilon(\sigma_f)=1$ equations numerically and find the $\sigma_f$ solution at the end of inflation for different $\Lambda$ values. 
We then plug this $\sigma_f$ into the Eq.~\eqref{eq:Ne} to find the start of inflation, i.e. $\sigma_i$, assuming inflation has lasted for 60 $e$-folds.

\begin{figure}[h!]
\centering
\includegraphics[height=80mm]{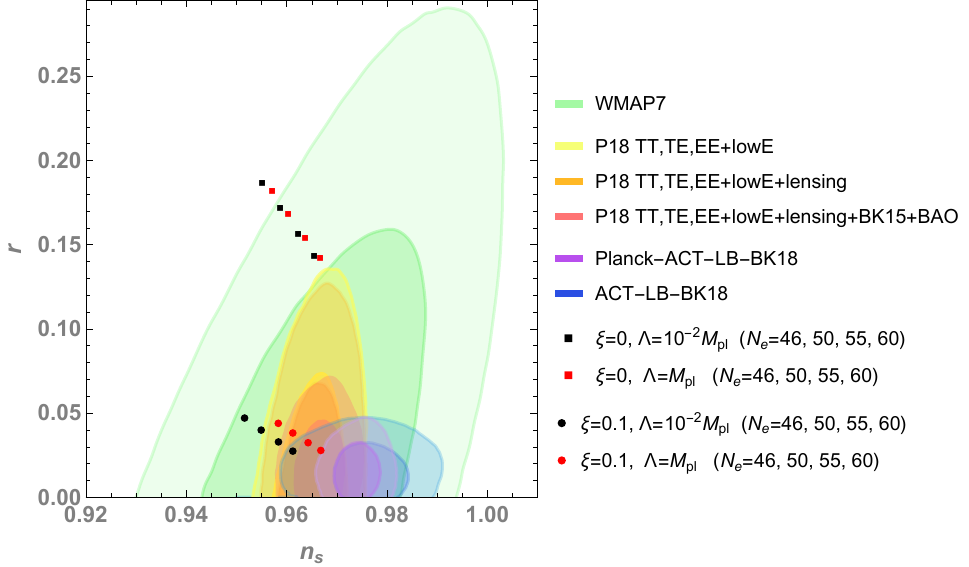}
\vspace{-2mm}
\caption{\label{fig:RIDE-plot} The predictions of the original RIDE model (square points) and the RIDE model with a non-minimal coupling to gravity (circle points) for the spectral index $n_s$ and tensor to scalar ratio $r$ as compared to the WMAP 7-year, Planck 2018 and ACT-BK18  data (the lighter and darker shades refer to the $95\%$ and $68\%$ confidence level regions, respectively). The red (black) points are for $\Lambda=\Mpl$ ($\Lambda=0.01\, \Mpl$) for values of $N_e=46$-$60$.}
\end{figure}

Next we calculate the amplitude of the scalar power spectrum, $A_s$, the tensor to scalar ratio, $r$, and the scalar spectral index, $n_s$ are given by 
\bea
A_s & = & \frac{1}{24\pi^2} \frac{1}{\epsilon_V} \frac{V}{\Mpl^4}  \,,  
\nonumber \\
r   & = & 16 \epsilon_V \,, \nonumber \\
n_s & = & 1-6\epsilon_V+2\eta_V \,. \label{eq:A_r_n}
\eea
The Planck 2018 limits are~\cite{Planck:2018jri}
\bea 
\ln(10^{10} \, A_s) &=& 3.044\pm 0.014 \\
n_s &=& 0.9649\pm 0.0042 \\
r &< & 0.06
\eea
For an estimate, we take the central vale of $A_s=2.1\times 10^{-9}$. The predictions of the original RIDE model for the spectral index $n_s$ and tensor to scalar ratio $r$ are within $95\%$ confidence level agreement with the WMAP 7-year  data. However, faced with the Planck 2018 data, the original RIDE model is not a viable inflationary model, as shown in Fig.~\ref{fig:RIDE-plot}. However, as discussed later in Section~\ref{sec:original-RIDE+gravity}, a similar analysis including a non-minimal coupling to gravity $\xi=0.1$ reduces $r$ and gives a good fit to Planck data, 
and this is also shown for comparison in Fig.~\ref{fig:RIDE-plot}.
The definition of data sets used in the analysis of the ACT collaboration are detailed in Appendix~\ref{sec:Appendix-B}.

\begin{figure}[ht!]
\centering
\includegraphics[width=0.48\linewidth]{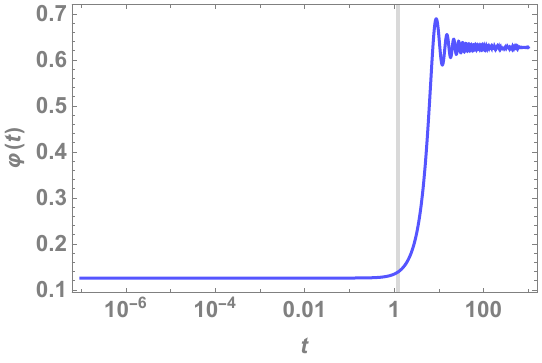}~~
\includegraphics[width=0.48\linewidth]{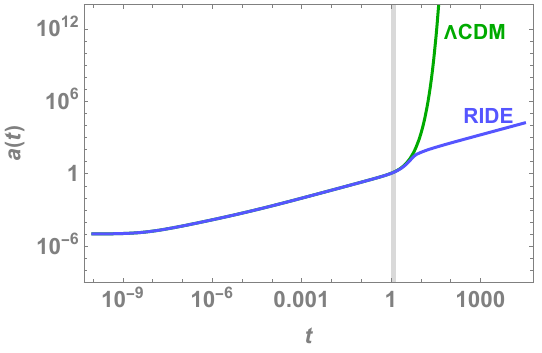}\\[2mm]
\includegraphics[width=0.48\linewidth]{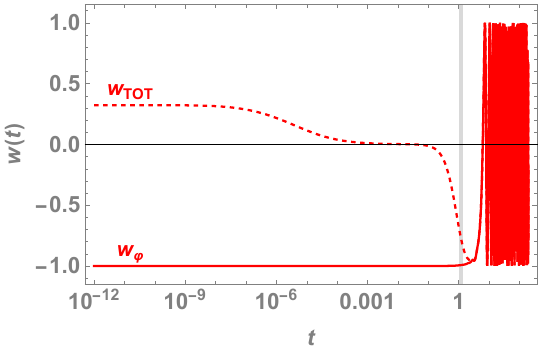}~~
\includegraphics[width=0.48\linewidth]{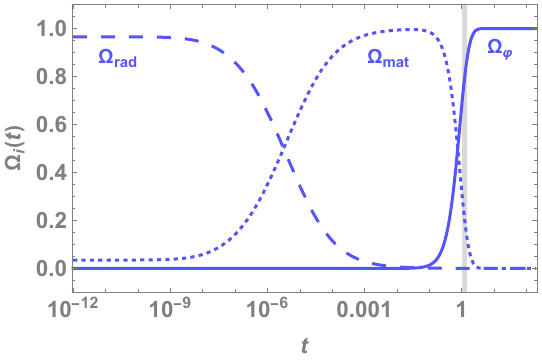}
\caption{The time evolution of the quintessence field $\varphi(t)$, scale factor $a(t)$, equation of state $w(t)$ and energy densities scaled by the critical density $\Omega_i(t)$, in the original RIDE model.
The time $t=1$ corresponds to the present day. }
\label{fig:Quintessence-field-RIDE}
\end{figure}
\subsection{Quintessence in the RIDE model}\label{sec:Quintessence}

In the RIDE model, the quintessence potential is as shown in Eq.~\eqref{eq:quint_pot-new}
with $v_\sigma = \langle \sigma \rangle =\sqrt{\frac{2}{e}}\, \Lambda$ as mentioned before.
Here dark energy is represented by the scalar field $\varphi$ that evolves in the potential $V(\varphi)$ as in Eq.~\eqref{eq:quint_pot-new}.
The energy density and pressure of the field come from its kinetic energy, $\dot{\varphi}^2/2$ and potential energy,  $V(\varphi)$.
The equation of state parameter, $w = p / \rho$, is not fixed like in $\Lambda$CDM ($w = -1$) but varies with time.

The dynamics of the quintessence field are governed by the Klein-Gordon equation in an expanding universe:
\be 
\ddot{\varphi} + 3H\dot{\varphi} + V'(\varphi) = 0,
\ee 
where $H$ is the Hubble parameter and $V'(\varphi) = dV/d\varphi$ represents the slope of the potential.
The energy density and pressure are:
\be 
\rho_{\varphi} = \frac{1}{2} \dot{\varphi}^2 + V(\varphi),
\ee 
\be 
p_{\varphi} = \frac{1}{2} \dot{\varphi}^2 - V(\varphi).
\ee
The equation of state parameter is:
\be 
w_{\varphi} = \frac{\dot{\varphi}^2 - 2V(\varphi)}{\dot{\varphi}^2 + 2V(\varphi)},
\ee 
which varies over time depending on the field dynamics as shown in Fig.~\ref{fig:Quintessence-field-RIDE}. There is no appreciable difference between the RIDE model predictions for dark energy and that of the cosmological constant as in the standard $\Lambda$CDM model up to the present day, although the 
future fate of the universe is predicted to be quite different.

\begin{figure}[ht!]
\centering
\includegraphics[height=90mm]{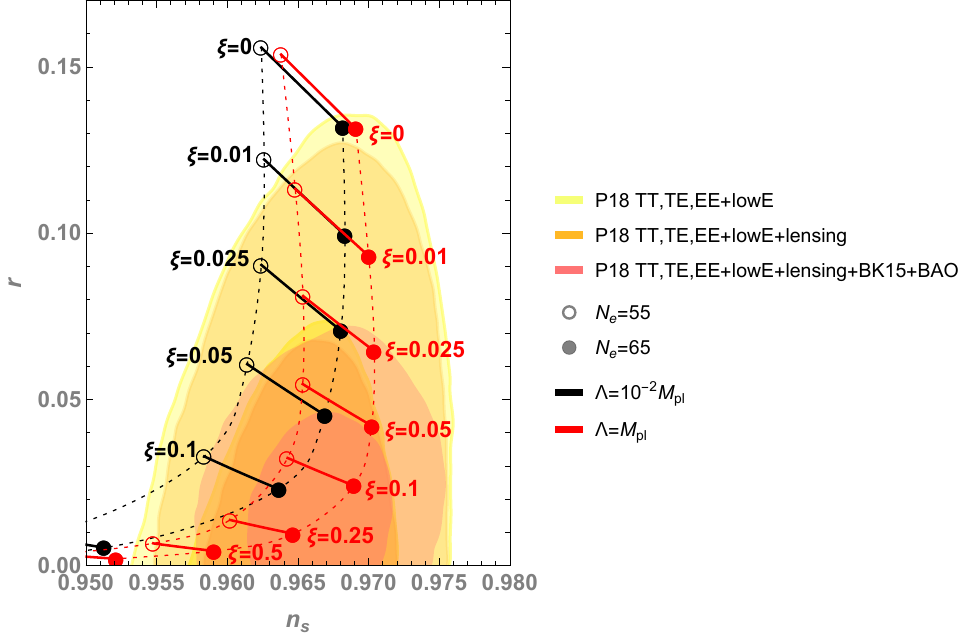}
\vspace{-4mm}
\caption{\label{fig:r-ns-plot-3} The $n_s$-$r$ plot for various $\xi$ values as compared to Planck 2018 data (with the zoomed in view only showing the most constraining Planck 2018 data near the bottom).
Two values of $\Lambda=\Mpl$ (red points) and 
$\Lambda = 0.01 \Mpl$ (black points) are considered.}
\end{figure}

\section{\label{sec:improved-RIDE}The RIDE model with extra couplings}

\subsection{\label{sec:original-RIDE+gravity} The RIDE model with a non-minimal coupling to gravity}

Thus far, we have considered the original RIDE potential as in Eq.~\eqref{eq:inf_pot-new}.
Henceforth, we allow the complex field $\Phi  =\frac{1}{\sqrt{2}}\, \sigma \, e^{i\varphi/v_\sigma}$ with $v_\sigma=\langle \sigma \rangle$, to couple to gravity.
The action of the model in the Jordan frame is:
\be 
\label{Eq:action-Jordan}
S_J  = \int d^4x \sqrt{-g}
\biggl[  \frac{1}{2} \Mpl^2 \, R 
+ \xi \, |\Phi|^2 \,R
- D_\mu \Phi^*  D^\mu \Phi
- V(\Phi)
\bigg],
\ee 
where $R$ is the Ricci scalar and the parameters $\xi$ is the dimensionless coupling of the $\Phi$ field to gravity. 
Expanding the action to show the explicit dependence on the inflaton field $\sigma$:
\be 
\label{eq:expanded-action-J}
S_J =
 \int d^4x \sqrt{-g}
\biggl[  \frac{\Mpl^2}{2}   \underbrace{\left(1 + \xi  \, \frac{\sigma^2}{\Mpl^2} \right)}_{\Omega^2}R \,
- \frac{g^{\mu\nu}}{2}  \underbrace{\left( \partial_\mu \sigma \partial_\nu \sigma + \frac{\sigma^2}{{v_\sigma}^2}\partial_\mu \varphi \partial_\nu \varphi \right)}_{\mbox{kinetic terms}}
-\underbrace{\frac{M^2}{2} \sigma^2  \ln \left( \frac{\sigma^2}{2\Lambda^2} \right)}_{V_J}\bigg]\,,
\ee 
where $\Omega^2$ is the conformal factor.
We show the details of the conformal transformation to the Einstein frame in the Appendix~\ref{sec:Appendix-A}.

For the purpose of our discussion below and presentation of our results, let us keep working with the inflaton field in the Jordan frame, i.e. $\sigma$, and write the potential in the Einstein frame as
\be  
V_E \, = \, \frac{V_J}{\Omega^4} = \frac{1}{2}\frac{M^2 \, {\sigma}^2}{\Omega^4}    \ln \left( \frac{{\sigma}^2}{2\Lambda^2} \right) 
= \frac{1}{2} M^2 \, {\sigma}^2 \left(1 + \xi  \, \frac{\sigma^2}{\Mpl^2} \right)^{-2} \ln \left( \frac{{\sigma}^2}{2\Lambda^2} \right) 
\label{VE1}
\ee 

Using the potential in Eq.\eqref{VE1}, we now repeat the inflation analysis in Section~\ref{sec:Inflation}, for various non-minimal couplings to gravity parameterised by $\xi$. Good fits to the Planck data are obtained for values of $\xi=0.1$, as shown earlier in 
Fig.~\ref{fig:RIDE-plot}. The general behaviour of the predictions in the $n_s$-$r$ plane as a function of $\xi$ is shown in 
Fig.~\ref{fig:r-ns-plot-3}.

\subsection{\label{sec:original-RIDE+quartic}The RIDE model with a quartic coupling} 

We now consider the additional effect of a quartic coupling, 
$\lambda \, |\Phi|^4$, leading to the following potential, 
\be 
\label{eq:pot-new}
V \approx   M^2 |\Phi|^2 \,  \ln \left( \frac{|\Phi|^2}{\Lambda^2} \right) + \,\lambda \, |\Phi|^4.
\ee 
The inflationary potential then becomes,
\be 
\label{eq:inf_pot-quartic}
V =  \frac{1}{2}M^2 \sigma^2  \ln \left( \frac{\sigma^2}{2\Lambda^2} \right)\, + \, \frac{1}{4} \lambda \, \sigma^4
\ee

\begin{figure}[h!]
\centering
\includegraphics[height=48mm]{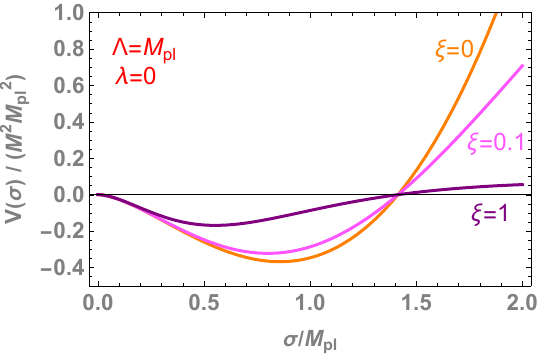}~~~~~~
\includegraphics[height=48mm]{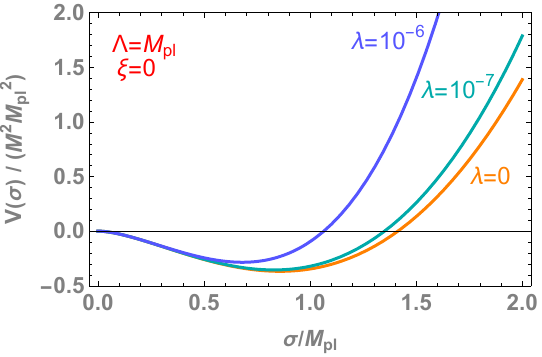}
\vspace{-3mm}
\caption{\label{fig:potentials-new}The shape of the inflationary potential for the RIDE model with a non-minimal coupling to gravity for varying $\xi$ couplings (left panel) and the RIDE model with a quartic term for varying $\lambda$ coupling (right panel) for $\Lambda=\Mpl$, for a typical value of $M \simeq (10^{-4}-10^{-3})\Mpl$.}
\end{figure}
The RIDE inflationary potential with the quartic and non-minimal coupling then becomes:
\footnote{The quintessence potential $V(\varphi)$ is unaffected by the introduction of the $\lambda$ quartic coupling or the non-minimal coupling to gravity, so the previous results discussed in Sec.~\ref{sec:Quintessence} are unchanged.}
\be  
\label{eq:Einstein-frame-pot-main}
V_E =  \biggl[ \frac{1}{2} M^2 \, {\sigma}^2  \ln \left( \frac{{\sigma}^2}{2\Lambda^2} \right) + \frac{1}{4} \, \lambda \, \sigma^4 \biggr] \left(1 + \xi  \, \frac{\sigma^2}{\Mpl^2} \right)^{-2}\,.
\ee   
The resulting potential is shown in Fig.~\ref{fig:potentials-new} for a different values of $\xi$ with $\lambda=0$ (left panel)
and different values of $\lambda$ with $\xi=0$ (right panel), for fixed $\Lambda=M_{\rm pl}$.
The effect of increasing $\xi$ (with $\lambda=0$) is both to flatten and distort the potential, leading to the results shown previously in Fig.~\ref{fig:r-ns-plot-3}.
Conversely, the effect of increasing $\lambda$ (with $\xi=0$) is to steepen the potential significantly, even for very small values of $\lambda$.
The original RIDE model potential (orange curve) with 
$\xi = \lambda=0$ is common to both panels in 
Fig.~\ref{fig:potentials-new}. 

\begin{figure}[ht!]
\centering
\includegraphics[height=90mm]{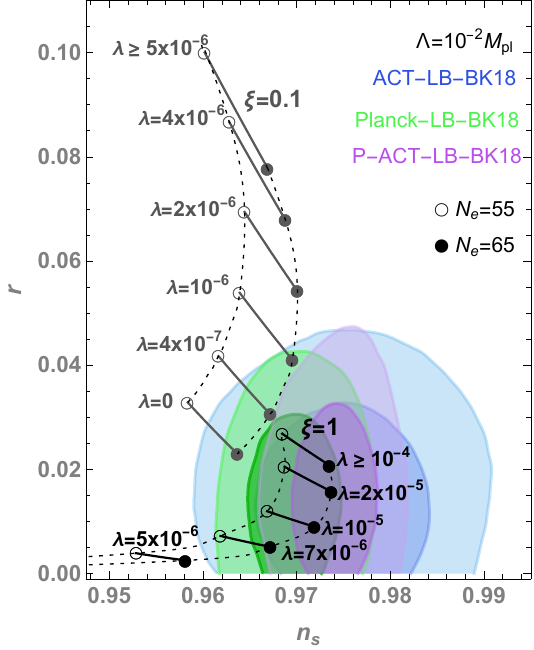}~~~~
\includegraphics[height=90mm]{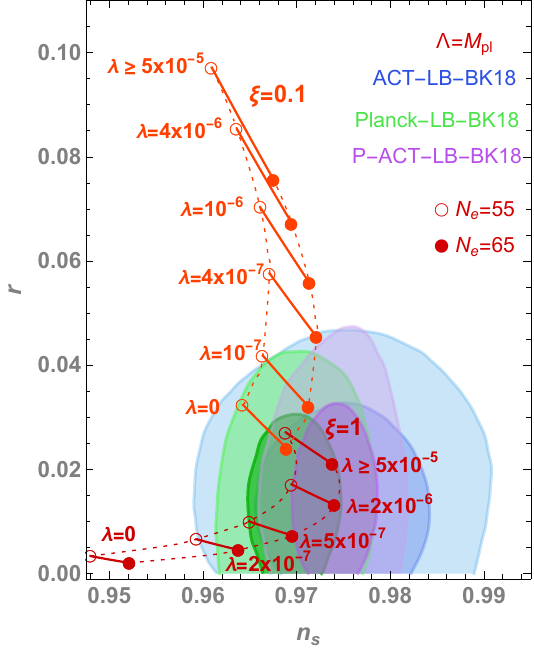}
\vspace{-4mm}
\caption{\label{fig:r-ns-plot-56} The predictions for the RIDE model in the $n_s$-$r$ plane, including a non-minimal coupling to gravity $\xi$ and a quartic coupling $\lambda$, as compared to the recent Planck and ACT data combined with LB-BK18 (see legends and Appendix~\ref{sec:Appendix-B} for details). The left (right) panel is for $\Lambda=10^{-2}M_{\rm{pl}}$ ($\Lambda=M_{\rm{pl}}$). In each panel, a ladder of predictions is shown for each of $\xi=0.1$ and $\xi=1.0$, with the rungs of the ladder corresponding to increasing values of $\lambda$, for two different values of the number of e-folds $N_e$. The value of $\xi=1.0$, for small values of $\lambda \sim 10^{-5}$, is preferred by the ACT data sets.}
\end{figure}

Allowing both non-zero $\xi$ and $\lambda$ at the same time, thus leads to a complex and subtle change in the shape of the potential, leading to the predictions shown in Fig.~\ref{fig:r-ns-plot-56} for two different values of $\Lambda=M_{\rm pl}$ (right panel) and 
$\Lambda=0.01 M_{\rm pl}$ (left panel), as compared to recent Planck and ACT data. 
To increase the spectral index $n_s$ prediction, as suggested by ACT results, it is necessary to increase the gravitational coupling, for example to $\xi=1$, in the presence of a small quartic coupling $\lambda$.

\section{\label{sec:Conclusions}Conclusions}

We have studied the simplest form of inflation, namely chaotic inflation with a quadratic potential $M^2 |\Phi|^2$, but generalized to include radiative corrections as parametrized by phenomenological potential 
$M^2 |\Phi|^2 \ln \left( |\Phi|^2/\Lambda^2 \right)$.
This radiatively corrected potential allows both inflation and dark energy to emerge from a single complex gauge singlet scalar field $\Phi$, and so is given the name RIDE (radiative inflation and dark energy). 
The RIDE potential has the classic Mexican hat shape which gives rise to spontaneous symmetry breaking at its minimum,
and hence a PNGB which can act as a quintessence field controlling dark energy. Thus inflation arises from the radial component of $\Phi$, while dark energy arises from its angular component, thereby unifying inflation and dark energy within a single field. 

Unfortunately, although the original RIDE model was consistent with the WMAP 7-year data, it is in tension with the more recent Planck 2018 and ACT observations. We have introduced additional couplings to the RIDE model in order to achieve consistency with current data. We have firstly considered a non-minimal coupling to gravity $\xi  |\Phi|^2 R^2$, 
which allows a good fit to Planck data
for $\xi\sim 0.1$, decreasing the tensor to scalar ratio to 
$r \lesssim0.03$. 
We have also studied the effect of adding 
a small additional quartic coupling $\lambda |\Phi|^4$, 
which could in principle also be radiatively generated.
By performing a detailed analysis of the available parameter space involving $\xi$ and $\lambda$, 
we have shown that small values of $\lambda \sim 10^{-5}$ 
together with $\xi\sim 1$
can serve to increase the value of the spectral index, while 
maintaining $r \lesssim0.03$, as preferred by the recent ACT and BK18 experiments.
Since both of these new couplings 
$\xi  |\Phi|^2 R^2$ and $\lambda |\Phi|^4$ 
only depend on the radial field 
$|\Phi|$, the behaviour of the axial quintessence field 
is therefore independent of 
$\xi$ and $\lambda$, and hence the dark energy predictions remain consistent with the standard $\Lambda$CDM cosmological model, as in the original RIDE model.

Finally we emphasise that recent studies introduce related modifications to various inflationary potentials to accommodate the ACT data, with somewhat similar results. 
In the present paper, these corrections have been applied to one of the simplest and earliest inflationary potentials, namely that of chaotic inflation, including radiative corrections leading to dark energy.

\subsubsection*{Acknowledgements}
VK acknowledges financial support from the Research Ireland Awards Grant 21/PATH-S/9475 (MOREHIGGS) under the SFI-IRC Pathway Program, and thanks the CERN-TH department for their hospitality. 
SFK thanks IFIC, Valencia, for hospitality and acknowledges the STFC Consolidated Grant ST/X000583/1; his work is funded by a Leverhulme Emeritus Fellowship Grant.

\appendix

\section{\label{sec:Appendix-A}The conformal transformation}

Here, we show the details of the conformal transformation from the Jordan frame, whose quantities are denoted without a tilde, to the Einstein frame in which quantities are denoted by a tilde, namely
\be
\label{eq:transformation-params}
\sqrt{-g} = \frac{1}{\Omega^4}\,  \sqrt{-\gt}\,, \qquad 
g^{\mu\nu} = \Omega^2 \, \gt^{\mu\nu}\,, \qquad 
R = \Omega^2\, \biggl( \Rt -\frac{3}{2}\,  \gt^{\mu\nu}\,  \partial_\mu \At \, \partial_\nu \At  \biggr)\,,
\ee
where we have introduced a new parameter $\At$ defined as $\At = \sqrt{\frac{2}{3}}\, \frac{A}{\Mpl}  = \ln \Omega^2$. The conformal factor can now be written as
\be 
\Omega^2  = 1 + \xi \left(\frac{\sigma}{\Mpl} \right)^2 = \, e^{\At} \,,
\ee 
allowing us to define the inflaton field $\sigma$ in the Jordan frame, in terms of $\At$, 
\be 
\left(\frac{\sigma}{\Mpl} \right)^2 = \frac{1}{\xi} \left(e^{\At}-1 \right) \,,
\ee
which will turn out to be the reparametrised inflaton field in the Einstein frame.

Let us now write the action in the Einstein frame 
\bea
S_E 
&=&
 \int d^4x \, \sqrt{-\gt} \,
\biggl[\frac{\Mpl^2 }{2}  \, \biggl( \frac{\Omega^2 \, R}{\Omega^4} \biggr) \,
- \frac{1}{2} \, \biggl( \frac{ g^{\mu\nu}}{\Omega^4}\biggr) \, \biggl( \mbox{kinetic terms} \biggr)
- \frac{V}{\Omega^4}
\bigg]
\nonumber\\
&=&
 \int d^4x \, \sqrt{-\gt} \,
\biggl[\frac{\Mpl^2 }{2}  \,\biggl(  \Rt -\frac{3}{2}\,  \gt^{\mu\nu}\,  \partial_\mu \At \, \partial_\nu \At\biggr)\,
- \frac{1}{2} \, \biggl( \frac{ \Omega^2 \, \gt^{\mu\nu}}{\Omega^4}\biggr) \, \biggl( \mbox{kinetic terms} \biggr)
- \frac{V}{\Omega^4}
\bigg] \nonumber\\
&=&
 \int d^4x \, \sqrt{-\gt} \,
\biggl[\frac{\Mpl^2}{2}\,\Rt 
- \frac{1}{2} \,\gt^{\mu\nu} \, \biggl(\partial_\mu A \, \partial_\nu A + \left(\frac{1}{\Omega^2}\right)\mbox{kinetic terms} \biggr)
- \frac{V}{\Omega^4}
\bigg]
\nonumber\\
&=&
 \int d^4x \, \sqrt{-\gt} \,
\biggl[\frac{\Mpl^2}{2}\,\Rt 
- \frac{1}{2} \,\gt^{\mu\nu}\, \underbrace{\biggl(\partial_\mu A \, \partial_\nu A + \left(\frac{1}{\Omega^2}\right)\mbox{kinetic terms} \biggr)}_{\mbox{new kinetic terms}}
- \underbrace{\frac{V}{\Omega^4}}_{\Vt}
\bigg],
\label{eq:action-E}
\eea
with the ``kinetic terms'' are as defined in Eq.~\eqref{eq:expanded-action-J} and 
\be 
\mbox{new kinetic terms} = 
\partial_\mu A \, \partial_\nu A
\, + \left(\frac{1}{\Omega^2}\right) 
\partial_\mu \sigma \partial_\nu \sigma 
\, + \left(\frac{1}{\Omega^2}\right)\, 
\frac{\sigma^2}{{v_\sigma}^2}\, \partial_\mu \varphi \partial_\nu \varphi \,,
\label{eq:new-kinetic}
\ee 
and need to be re-written in a canonical form:
\be  
\biggl(\frac{1}{\Omega^2}\biggr)\partial_\mu \sigma \partial_\nu \sigma 
= \frac{1}{6\, \xi} \, \left( \frac{1}{1- e^{-\At}} \right) \partial_\mu A \,\partial_\nu A \,.
\ee 
Recall that $\varphi$ does not contribute to the inflation process, as discussed in Sec.~\ref{sec:original-RIDE}, and is not a dynamical field during inflation.
Therefore, the kinetic terms in the Einstein frame in Eq.~\eqref{eq:new-kinetic} reduce to
\be 
\mbox{new kinetic terms} = 
\left[ 1+ \frac{1}{6\, \xi} \left(\frac{1}{1-e^{-\At}} \right)    \right]
\partial_\mu A \, \partial_\nu A \, ,
\ee 
which is canonical in the usual $6\xi \gg 1$ limit~\cite{Keus:2024khd,Rubio:2018ogq}.
The potential in the Einstein frame in Eq.~\eqref{eq:Einstein-frame-pot-main} can be written as
\bea 
\Vt \, = \, \frac{V}{\Omega^4} &=& 
\biggl[ \frac{1}{2} M^2 \, {\sigma}^2  \ln \left( \frac{{\sigma}^2}{2\Lambda^2} \right) + \frac{1}{4} \, \lambda \, \sigma^4 \biggr] \left(1 + \xi  \, \frac{\sigma^2}{\Mpl^2} \right)^{-2}
\nonumber 
\\
&=& 
\frac{M^2}{2} \left(\frac{\Mpl^2 }{\xi}\right)  \left(\frac{e^{\At}-1}{e^{2\At}} \right)
 \ln \left[ \left(\frac{\Mpl^2 }{\xi}\right) \frac{e^{\At}-1 }{2 \,\Lambda^2} \right]
+ 
\frac{\lambda}{4}  \left(\frac{\Mpl^2 }{\xi}\right)^2 \left(1- e^{-\At} \right)^2 .\,\,\,
\eea

\section{\label{sec:Appendix-B}Datasets definitions in the plots}

For completeness, we show the definition of data sets used in the analysis of the ACT collaboration~\cite{ACT:2025tim} which appear in our plots.

\begin{itemize}

\item \textit{Planck}: Low-$\ell$ and full-sky CMB spectra from Planck's final public release (PR4), which includes TT, TE, EE with updated Sroll2 low-$\ell$ likelihood.

\item \textit{TT}: Temperature-temperature power spectra.

\item \textit{TE}: Temperature-E mode polarization power spectra.

\item \textit{EE}: Polarization-polarization power spectra.

\item \textit{Sroll}: A map-making algorithm and likelihood framework designed to improve the analysis of large angular scales (low multipoles $\ell < 30$) CMB polarisation data from the Planck High Frequency Instrument.
Sroll1 was introduced to address instrumental systematics that limited the accuracy of earlier Planck low-$\ell$ polarisation measurements. 
An improved version, Sroll2 (low-$\ell$ Planck likelihood, providing large-scale $E$-mode information), includes better calibration, de-striping, and foreground modelling, leading to tighter constraints on parameters such as the reionisation optical depth $\tau$. Low-$\ell$ $E$-mode polarisation is crucial for constraining the optical depth to reionisation $\tau$, which in turn affects the amplitude of the scalar perturbations $A_s$ and indirectly the inferred value of the Hubble constant $H_0$.
The Sroll likelihoods provide an independent and more robust handle on these large-scale modes compared to previous Planck analyses.

\item \textit{CMB Lensing}: Power spectra lensing reconstruction likelihoods used to break degeneracies in cosmological parameters.

\item \textit{BK15 (BK18)}: BICEP/Keck 2015 (2018) $B$-mode polarisation likelihood, used to constrain the primordial tensor-to-scalar ratio $r$.

\item \textit{BAO}: Baryon Acoustic Oscillation data from DESI Year~1, a collection of 12 measurements from galaxy, quasar, and Ly$\alpha$ tracers spanning $0.1 < z < 4.2$, occasionally replaced by Baryon Oscillation Spectroscopic Survey (BOSS) BOSS/eBOSS for robustness tests.
To ensure that the data is not solely driven by DESI, and in light of some $2.5\sigma$ deviations between the DESI luminous red galaxy (LRG) data points and previous measurements at the same redshifts, analyses with DESI replaced by BOSS/eBOSS BAO data, including both BOSS DR12 LRGs and eBOSS DR16 LRGs is also considered, used usually for models in which BAO data have a significant impact on the parameter constraints.

\item \textit{ACT}: High-$\ell$ temperature and polarisation spectra and high-resolution CMB power spectra from the Atacama Cosmology Telescope (ACT) DR6 which includes TT, TE, EE spectra and Sroll2 (low-$\ell$) likelihood.

\item \textit{LB}: Adding CMB lensing and BAO.
The ACT-LB-BK18 dataset combination is usually used when investigating primordial gravitational waves (models with a free tensor-to-scalar ratio $r$). Including BK18 allows a joint constraint on $r$ alongside the scalar sector parameters, strengthening bounds on $r$ beyond what ACT or Planck can achieve alone.

\item \textit{P-ACT}: The combination of ACT DR6 and Planck primary anisotropies.
The P-ACT-LB combination is used as the default ``baseline'' data combination, since it provides the tightest constraints in most extended models. 
The joint best-fit $\Lambda$CDM model, P-ACT-LBS, which also includes SNIa Pantheon+ is an excellent fit, shown to improve over Planck-alone or ACT-alone constraints due to complementarity in multipole coverage.

\item \textit{SNIa Pantheon+}: 1550 spectroscopically confirmed SNe Ia from 18 subsamples, $0.001 < z < 2.26$, used when exploring models that affect the late-time expansion.

\end{itemize}

\bibliographystyle{JHEP} 
\bibliography{RIDE-again.bib}

\end{document}